\documentclass[reprint,amsmath,amssymb,prc]{revtex4-1}
\usepackage[utf8]{inputenc}
\usepackage{color}
\usepackage{graphicx}

\newcommand{\AlO}{\ensuremath{{\text{Al}_2\text{O}_3}}}
\newcommand{\SiO}{\ensuremath{\text{SiO}_2}}
\newcommand{\Sp}{\ensuremath{S_{21}}}
\newcommand{\Sps}{\ensuremath{\left|\Sp\right|^2}}
\newcommand{\Lq}{\ensuremath{\lambda/4}}
\newcommand{\eeff}{\ensuremath{\epsilon_\text{eff}}}
\newcommand{\un}[1]{\ensuremath{\,\text{#1}}}

\newcommand{\fr}{\ensuremath{f_{\text{r}}}} 
\newcommand{\Qi}{\ensuremath{Q_{\text{i}}}} 
\newcommand{\Qe}{\ensuremath{Q_{\text{e}}}} 
\newcommand{\Qc}{\ensuremath{Q_{\text{c}}}} 
\newcommand{\Ql}{\ensuremath{Q_{\text{l}}}} 

\begin{document}

\title{Towards carbon nanotube growth into superconducting microwave resonator 
geometries}

\author{S. Blien}
\author{K. J. G. G\"{o}tz}
\author{P. L. Stiller}
\author{T. Mayer}
\author{T. Huber}
\author{O. Vavra}
\author{A. K. H\"{u}ttel}
\email[]{andreas.huettel@ur.de}
\affiliation{Institute for Experimental and Applied Physics, 
University of Regensburg, 93040 Regensburg, Germany}

\date{\today}

\begin{abstract}
The in-place growth of suspended carbon nanotubes facilitates the observation 
of both unperturbed electronic transport spectra and high-$Q$ vibrational 
modes. For complex structures integrating, e.g., superconducting rf elements 
on-chip, selection of a chemically and physically resistant material that 
survives the chemical vapor deposition (CVD) process provides a challenge. 
We demonstrate the implementation of molybdenum-rhenium coplanar waveguide 
resonators that exhibit clear resonant behaviour at cryogenic temperatures 
even after having been exposed to nanotube growth conditions. The properties 
of the MoRe devices before and after CVD are compared to a reference niobium 
device.
\end{abstract}

\maketitle

\section{Introduction}
The integration of different types of mesoscopic systems into hybrid device 
geometries has led to a multitude of experimental insights. One recent 
development is the application of coplanar waveguide resonator technology in 
hybrid quantum systems, see e.g. \cite{rmp-xiang:623} for a detailed review of 
performed and possible experiments. Recently significant advances were made 
combining superconducting rf systems with with semiconductor quantum dots 
\cite{prl-frey:046807,prb-basset:125312,nature-petersson:380}, and also in 
particular carbon nanotubes. There, published results by now target solid state 
cavity quantum electrodynamics \cite{prb-viennot:165404,ncomm-ranjan:7165} 
and manipulation of quantum states towards quantum information processing 
\cite{science-viennot:408}. 

On their own, clean carbon nanotubes, as quasi-one dimensional carbon 
macromolecules, excel in their electronic 
\cite{nmat-cao:745,nature-kuemmeth:448,nnano-steele:363,brokensu4,%
ncom-ranjan:7165,rmp-laird,fabryperot} as well as 
nanomechanical properties 
\cite{highq,nnano-moser:1007,strongcoupling,magdamping,heliumdamping,ilani2}.
This emerges in particular clearly if fabrication steps after nanotube growth, 
as, e.g., wet chemistry or lithography, are kept to a minimum.
Several strategies to that effect are possible. A  well established method is to
grow the carbon nanotubes via chemical vapor deposition (CVD) in situ across 
pre-defined electrodes \cite{nmat-cao:745}. This generally leads to low contact 
resistance as well as devices resilient to mechanical vibrations or temperature 
changes. At the same time, the choice of materials is strongly 
restricted; contacts and other chip structures as, e.g., gates or isolation 
oxides, have to survive the CVD process, where the devices are exposed to hot 
and chemically aggressive gas mixtures. Metal thin films melt or deform strongly,
or incorporate carbon or hydrogen. In particular, superconductors may display a
strong decrease of their critical temperature. In addition the carbon nanotubes grow at 
random orientation, lowering the device yield.

This has recently led to the development of alternative strategies. 
Typically the nanotubes are grown on a separate substrate and then transferred 
onto the readily structured device at last instance 
\cite{ilani2,nl-wu:1032,nnano-pei:630,ilani,pssb-gramich:2496}. The separation 
of growth and measurement allows a wider choice in device materials and 
targetted placement of the nanotubes, but may come at other costs. Surface 
oxidation or contamination of the metallic electrodes may require an additional 
annealing step to obtain transparent contacts; even cleaning the metal
surfaces in situ using argon ion-etching and then keeping them in vacuum during 
transfer and measurement \cite{ilani} has been performed.

Here, in preparation for the first, ``overgrowth'' approach, we use the CVD 
process environment to simulate nanotube growth across pre-defined electrodes 
and investigate metal films with/without having undergone CVD. In particular 
with regard  to high-frequency experiments, where a suspended carbon nanotube 
is to be coupled to, e.g., a coplanar microwave resonator, a physically and 
chemically stable metal is required that remains superconducting even after 
incorporation of carbon and possible segregation processes during the 
high-temperature step.
\begin{figure*}[t]%
\begin{center}
\includegraphics*[width=\textwidth]{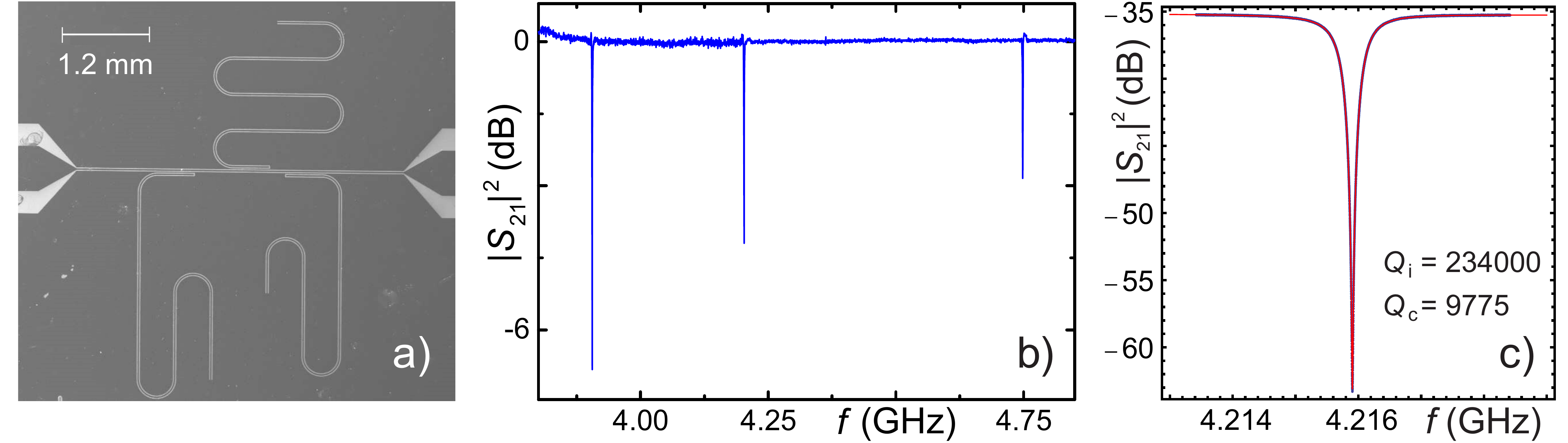}
\end{center}
\caption{\label{resintro} (Color online)
(a) SEM image of a device containing three \Lq\ superconducting 
coplanar waveguide resonators of different length, coupled to a common feed 
line, on a dielectric substrate. (b) Transmission spectrum ($T=4.2\un{K}$) of 
the feed line of a device with geometry as depicted in (a); 130nm niobium 
sputtered onto a silicon substrate with 500nm thermally grown \SiO\ cap layer. 
The spectrum is normalized to the high-power transmission where 
superconductivity within the resonators is suppressed. (c) Exemplary detail 
measurement of a resonance of the device from (b) at $T=15\un{mK}$; the solid 
red line is a fit (see text) resulting in $\Qi=234000$ and $\Qc=9775$.}
\end{figure*}
We characterize coplanar waveguide resonators fabricated from the co-sputtered 
molybdenum-rhenium alloy $\text{Mo}_{20}\text{Re}_{80}$ \cite{remo}. Even after 
CVD clear resonant behaviour is observed. We compare the device parameters with 
a reference niobium device and analyze the temperature dependence of resonance 
frequency and quality factor. 

\section{Carbon nanotube CVD growth process}

Our carbon nanotube growth via CVD follows a well-established process 
\cite{nature-kong:878} that is able to produce few clean single-wall carbon 
nanotubes. The nanotubes grow from well-defined positions and subsequently fall 
over contact structures.

A catalyst suspension consisting of iron(III) nitrate nonahydrate
$\text{Fe}(\text{NO}_3)_3\cdot 9\text{H}_2\text{O}$, molybdenum 
dioxydiacetylacetonate $\text{MoO}_2(\text{acac})_2$, and 
$\text{Al}_2\text{O}_3$ nanoparticles in methanol is used. The catalyst 
deposition area is defined lithographically. Following development of the 
electron-beam resist, the catalyst suspension is drop-cast onto the chip, and 
the solvent is evaporated on a hot plate at 150$\,^\circ \text{C}$. After a 
lift-off step, only the catalyst in the growth area remains on the chip. The 
devices are then placed into a $1''$ quartz tube, where they are heated up 
in an argon and hydrogen gas flow to a temperature of $850\,^\circ \text{C}$. 
As soon as the temperature is reached, the argon flow is stopped and the 
devices are exposed to a hydrogen ($20\un{sccm}$) / methane ($10\un{sccm}$) 
atmosphere for ten minutes. During this time, the growth of the carbon nanotubes 
takes place. Afterwards, the devices are cooled down again under argon and 
hydrogen flow. Nanotube devices can now be electronically tested and 
pre-characterized. 

For the thin-film material tests presented here, focussing on the effect of the 
high-temperature process on the metal films of the devices, both lithography 
and catalyst deposition steps are omitted for simplicity. The high-temperature 
process is performed in an identical way as actual nanotube CVD growth.

\section{Resonator device fabrication and basic properties}

Two types of substrate form the starting point of device fabrication, either  
crystalline \AlO\ or compensation-doped Si with a $500\,\text{nm}$ thermally 
grown oxide. A metal layer is sputter-deposited and structured via optical 
lithography and reactive ion etching. Fig. \ref{resintro}(a) shows a resulting 
reference 
device, in this case made of niobium on a Si/\SiO\ substrate, containing three 
$\lambda/4$ resonators capacitively coupled to a common feed line. In 
Fig.~\ref{resintro}(b), in a transmission measurement at $T=4.2\,\text{K}$ the 
resonances of the $\lambda/4$ structures can be clearly identified as distinct 
minima of the feed line power transmission \Sps.

The required resonator lengths for intended design frequencies can be calculated
following 
\begin{equation}
\fr = \sqrt{\frac{1-\alpha}{\eeff}}
\frac{c}{4l}
\label{f-l}
\end{equation}
where $\alpha$ is the kinetic inductance fraction of the superconductor, 
discussed below in more detail, $l$ is the length of the resonator, and \eeff\ 
is an effective relative permittivity resulting from the substrate below and the 
vacuum (or helium, depending on the measurement; both with $\epsilon_r\simeq 
1$) above the substrate. In the simple case of a uniform substrate with 
permittivity $\epsilon_r$, \eeff\ can be approximated as $\eeff = 
(\epsilon_r+1)/2$.

Fig.~\ref{resintro}(c) displays a detail measurement of the reference device 
from (b) in a dilution refrigerator at $T=15\un{mK}$. The overall transmission 
values are setup-specific, resulting from an attenuation of $-53\un{dB}$ for 
thermal coupling in the signal input line and an amplification of $+29\un{dB}$ 
via a HEMT amplifier \cite{caltech} at the 1K stage in the signal output line. 
We experimentally determine a center frequency of $\fr\approx 4.2159\un{GHz}$.

Aside from the center frequency \fr, the resonance is characterized by the 
intrinsic quality factor \Qi, describing energy loss within the resonator, 
and the coupling quality factor \Qc, describing energy transfer from and to the 
feed line. The so-called loaded quality factor $\Ql=1/\left(1/\Qi+1/\Qc\right)$ 
combines both. The solid line in Fig.~\ref{resintro}(c) is a fit following 
\cite{jap-khalil:054510,apl-bruno:182601}, where the transmission 
(scattering matrix) parameter \Sp\ is 
expressed as 
\begin{equation}
S_{21}=1-\frac{\frac{\Ql}{|\Qe|}\; e^{i \theta}}{1+2 i \Ql 
\frac{f-\fr}{\fr}}.
\label{eq:S21}
\end{equation}
Here, $\Qe=|\Qe|\text{e}^{-i\theta}$ has been additionally introduced, 
a complex-valued parameter related to the coupling quality factor 
\Qc\ as $\Qc^{-1}=\text{Re}\,\Qe^{-1}$, while $\text{Im}\, \Qe$ represents 
resonance asymmetries.
For the reference device of Fig.~\ref{resintro}(c) we find at $T=15\un{mK}$ values of 
$\Qi =234000$ and $\Qc=9775$, demonstrating the quality of our niobium films 
and the lithographic patterning as well as the function of our measurement 
setup.

\begin{figure*}[t]
\begin{center}
\includegraphics[width=0.8\textwidth]{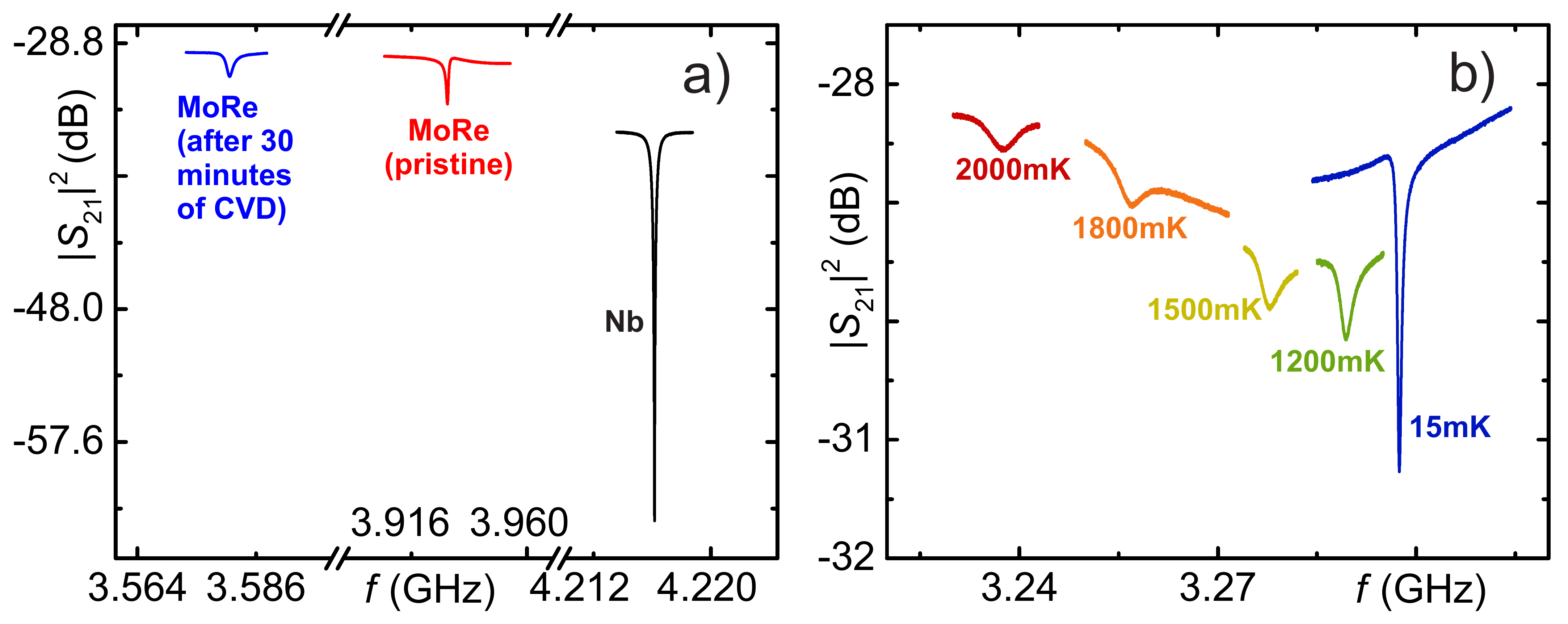}
\end{center}
\caption{\label{fitting}(Color online)
(a) Power transmission resonances $\Sps(f)$ at $T=15\un{mK}$ of three lithographically 
identical \Lq\ structures (see Fig.~\ref{resintro}(a) for the optical mask 
geometry). The patterned thin film consists of 130nm Nb (right resonance), 
145nm pristine Mo$_{20}$Re$_{80}$ (middle resonance), and 145nm 
Mo$_{20}$Re$_{80}$ after undergoing 30 min of CVD process (left resonance). The 
distinct shift in the resonance frequency \fr\ is caused by different kinetic 
inductance fractions $\alpha$ for the different materials. 
(b) $\Sps(f)$ at the resonance of the CVD-treated Mo$_{20}$Re$_{80}$
device from (a), for temperatures $15\un{mK}\le T \le 2\un{K}$.} 
\end{figure*}

\section{Molybdenum-rhenium as resonator material}
\label{sec-more}

While niobium is a well-established material for superconducting coplanar 
radiofrequency circuit elements, its thin films unfortunately do not survive 
the conditions of CVD carbon nanotube growth. Rhenium thin films are stable 
under CVD conditions \cite{brokensu4,magdamping}, however in our observation 
the critical temperature typically decreases below $1\un{K}$. A highly 
promising material is given by molybdenum-rhenium alloys. Pristine films have 
been shown to exhibit critical temperatures up to $15\un{K}$ 
\cite{ssc-testardi:1892,apl-gavaler:179,pssa-postnikov:21}. In addition, the 
films remain stable under CVD conditions 
\cite{remo,sr-schneider:599,apl-singh:222601}. 
While a significant amount of carbon is integrated into the metal, the high 
temperature process even leads to annealing-like processes and an initial 
increase in critical temperature, current and field \cite{remo,apl-singh:222601,apl-aziz:233102}.

Fig. \ref{fitting}(a) shows transmission resonances of three lithographically 
identical \Lq\ structures, using different metallization layers. The rightmost 
resonance at highest frequency (black) uses niobium, the middle one (red) 
pristine molybdenum-rhenium, and the left one (blue) molybdenum-rhenium which 
has undergone the CVD process used for carbon nanotube growth. The 
molybdenum-rhenium 20:80 thin films have been deposited via simultaneous 
sputtering from two sources and their composition, identical for all devices 
discussed here, has been verified via x-ray photoelectron spectroscopy 
\cite{remo}.

The difference in resonance frequency predominantly stems from a difference in 
the so-called kinetic inductance of the waveguide: the inertia-delayed response 
of charge carriers to a high-frequency field is mathematically equivalent to an 
increased inductance and can be described as such. It leads to an additional 
contribution $L_k$ to the inductance 
per length of the coplanar waveguide. Given the geometric, 
temperature-independent inductance per length $L_g$ of a waveguide, the 
so-called kinetic inductance fraction is then defined as $\alpha=L_k / 
(L_g+L_k)$. 

From the data of Fig.~\ref{fitting}(a), assuming a small kinetic inductance 
fraction for niobium $\alpha_\text{Nb}\simeq 0$ at dilution refrigerator base 
temperature $T=15\un{mK}$, we obtain using Eq. \ref{f-l} from the resonance 
frequency for the molybdenum-rhenium device before CVD 
$\alpha_\text{MoRe} = 1-(f_0^\text{MoRe}/f_0^\text{Nb})^2 = 0.131$ and 
after CVD $\alpha_\text{MoRe,CVD}=0.279$.

Via fitting to Eq.~\ref{eq:S21} we obtain the quality factors \Qi\ and \Qc\ of 
the resonator structures.
The coupling quality factor \Qc\ is similar ($Q_c\sim 10^4$) for all  three devices, 
consistent with the identical geometry and thereby coupling capacitance.
For the intrinsic quality factor we here obtain $\Qi=234000$ 
for Nb, $\Qi=20800$ for MoRe without CVD exposure, and $\Qi=2700$ for MoRe after 
30min CVD exposure. 
 The intrinsic quality factor differs strongly, due to the presence of
quasiparticle excitations and corresponding dissipation (see following section).

\section{Temperature dependence of resonator properties}

Fig.~\ref{fitting}(b) shows the resonance of the CVD-treated molybdenum-rhenium device from 
Fig.~\ref{fitting}(a), at differing temperature. The resonance 
frequency \fr\ decreases strongly with temperature, as does the total quality 
factor. A quantitative description of this is given by the so-called 
Mattis-Bardeen theory \cite{remo,pr-mattis:412}. The temperature dependence of 
\fr\ and \Qi\ can be expressed as
\begin{equation}\label{eq:freq-mb}
\frac{\fr-f_0}{f_0}  = \frac{\alpha_0}{2} \, \frac{\delta 
\sigma_{2}}{\sigma_{2}},
\end{equation}
\begin{equation}\label{eq:q-mb}
\delta\left( \frac{1}{\Qi}\right)=\alpha_0 \frac{\delta \sigma_1}{\sigma_2},
\end{equation}
with $f_0$ and $\alpha_0$ as the zero-temperature limit of resonance frequency 
and kinetic inductance fraction. $\sigma_1$ and $\sigma_2$ are the 
real and imaginary part of the complex conductivity $\sigma$ of the device. We 
use analytical approximations for both parts of the complex conductivity
in the low temperature limit \cite{remo,gao-zmuidzinas}, containing the temperature
dependent BCS energy gap and the normal state conductivity.

\begin{figure}[htb]%
\includegraphics*[width=\columnwidth]{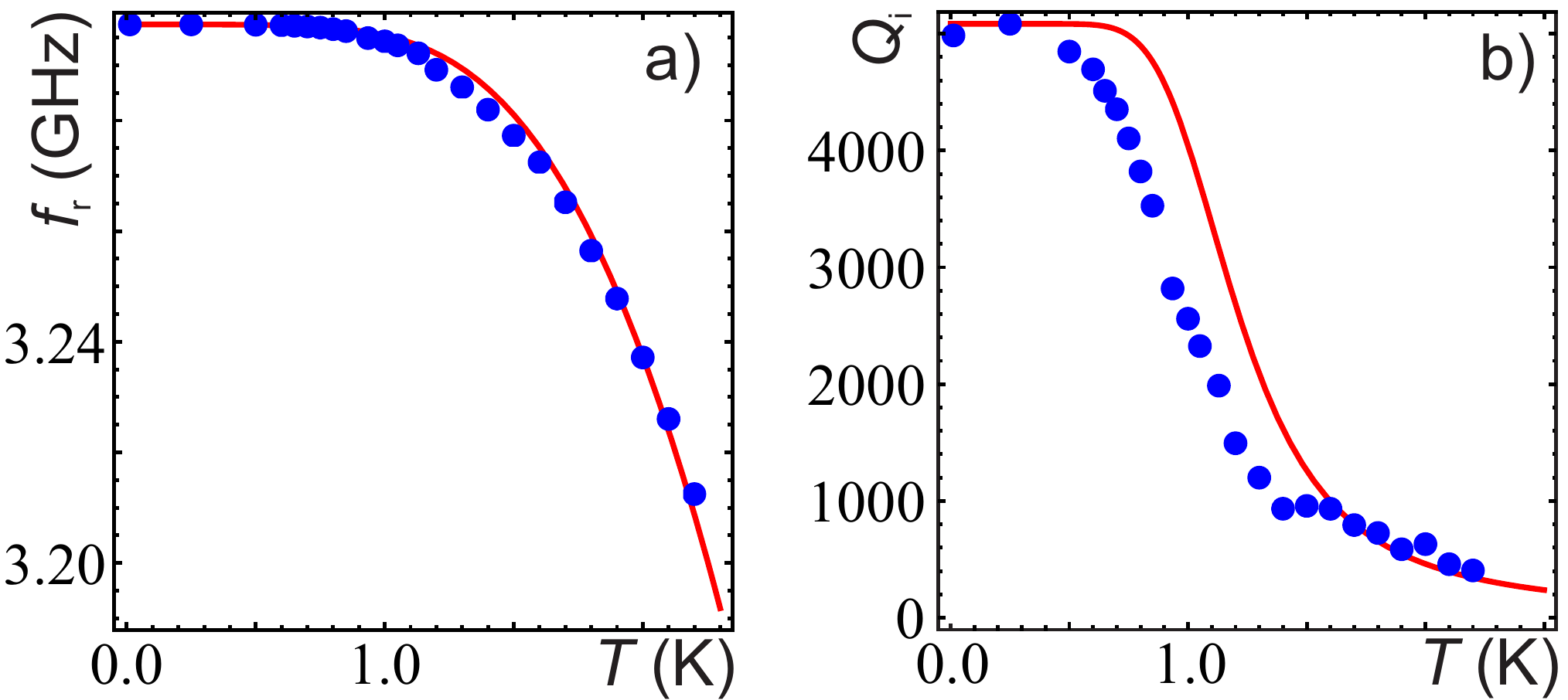}
\caption{\label{MB}
(a) Resonance frequency $\fr(T)$ and (b) internal quality factor $\Qi(T)$ as function of 
temperature, for the device of Fig. \ref{fitting}(b) ($145\un{nm}$ MoRe, after 
$30\un{min}$ CVD conditions). The red lines show fit curves using 
Mattis-Bardeen-theory, see the main text.
} 
\end{figure}

Figure~\ref{MB} shows
the extracted resonance frequency \fr\  (Fig.~\ref{MB}(a)) and internal quality factor 
\Qi\ (Fig.~\ref{MB}(b)) as function of temperature $T$. The solid line in Fig.~\ref{MB}(a)
is a fit following Eq.~\ref{eq:freq-mb}, with $\alpha_0$ as the only free fit parameter.
Using a critical temperature of $T_{\text{c}}=4\,\text{K}$ determined via dc measurements,
the fit results in a kinetic inductance fraction of $\alpha= 0.361$. This is larger than the 
value we found by comparing the resonance frequencies of different materials in 
Sec.~\ref{sec-more}, however, a smooth drop of the resistance in dc 
measurements makes a precise determination of $T_\text{c}$ difficult.

In the same manner, we obtain $\alpha_0=0.176$ for the pristine MoRe device, 
and $\alpha_0=0.029$ for the niobium device. The value of $\alpha_0$ for the 
MoRe device is comparable to the one of Sec.~\ref{sec-more}, while the value 
for the niobium device confirms that $\alpha \approx 0$ was a good 
approximation. Generally, we find small kinetic inductance fractions for 
niobium and intermediate ones for MoRe, with increasing values for increasing 
CVD exposure time, corresponding to deterioration of the superconducting 
properties.

Regarding the intrinsic quality factor \Qi\ in Fig.~\ref{MB}(b), we use the 
value $\alpha_0= 0.361$ extracted from the fit of Fig.~\ref{MB}(a), and plot 
the theoretical result following Eq.~\ref{eq:q-mb} without additional free 
parameters. The result (solid red line) displays acceptable agreement with the 
data points. However, models more complex than the straightforward 
Mattis-Bardeen description may be appropriate, see, e.g., \cite{zem} for the 
integration of a finite quasiparticle lifetime.

\begin{figure*}[htb]
\includegraphics[width=\textwidth]{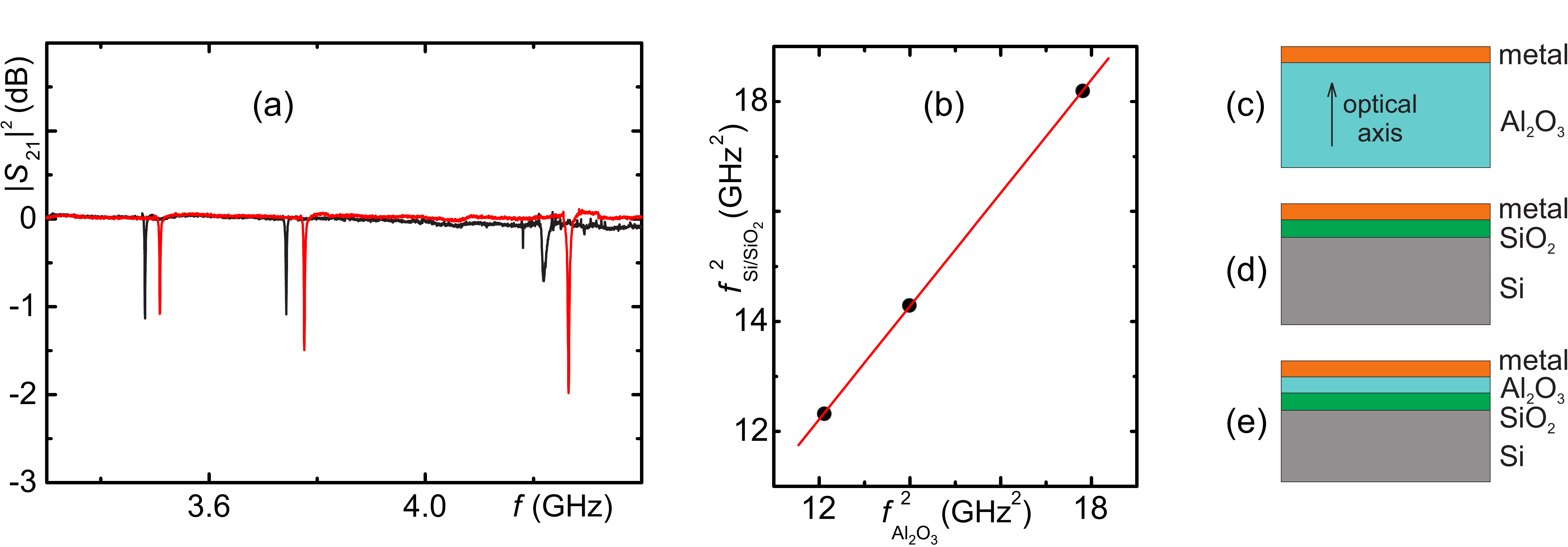}
\caption{
(a) Transmitted signal \Sps\ for two devices made of molybdenum-rhenium using 
the same optical mask, but different substrate materials. The shift in resonance 
frequency is caused by different relative permittivities of the substrates 
(red: $\text{Si/SiO}_2$, black: $\text{Al}_2\text{O}_3$). (b) The ratio of the 
squared resonance frequencies is used to determine the relative permittivity of 
$\text{Si/SiO}_2$. (c) - (e) Different substrate structures: (c) crystalline 
\AlO, (d) compensation-doped silicon with 500nm of \SiO\ on top, (e) same as in 
(d) with an additional layer of \AlO\ that was grown using atomic layer 
deposition.} 
\label{substrate}
\end{figure*}
\section{Impact of the substrate material}

Two different substrate materials have been tested, on one hand crystalline 
\AlO\ (Fig. \ref{substrate}(c)), on the other highly resistive silicon with a 
thermally (dry) grown, 500nm thick \SiO\ cap layer (Fig. \ref{substrate}(d)). 
The dielectric constant of the substrate $\epsilon_r$ enters the resonance 
frequency via Eq.~\ref{f-l}. Fig.~\ref{substrate}(a) displays the transmission 
spectrum of two devices using the same lithographic structure and metal film 
(pristine MoRe), but different substrates. 

As a consistency check, we can compare the resonance positions, deduce the 
effective permittivity of the Si/\SiO\ bilayer, and compare it with a 
calculation. Using Eq.~\ref{f-l} it is easy to show that the effective 
dielectric constants of the two devices relate to the resonance frequencies as 
\begin{equation}
\frac{\epsilon_{\text{eff}}^\AlO}{\epsilon_\text{eff}^{\text{Si}/\SiO}}= 
\left( \frac{\fr^{\text{Si}/\SiO}}{\fr^\AlO} \right)^2, 
\end{equation}
with $\epsilon_{r,\perp}^\AlO=9$ and $\epsilon_{\text{eff}}^\AlO=5$.
From the data of Fig.~\ref{substrate}(a,b) we obtain  
$\epsilon_\text{eff}^{\text{Si}/\SiO} = 4.86$ and thereby  
$\epsilon_r^{\text{Si}/\SiO} = 8.72$. Using the geometrical device dimensions 
and following \cite{chenandchou} leads to an expected value of 
$\epsilon_\text{eff}=4.88$ and thereby to excellent agreement with the 
experimentally determined value.

For integrating carbon nanotube quantum dot structures with coplanar 
resonators, additional gate isolation layers are desirable. With this in 
mind we have tested additionally inserted oxide layers below the 
superconducting coplanar waveguide. A sketch of this is shown in 
Fig.~\ref{substrate}(e), where an additional \AlO\ layer has been grown using 
atomic layer deposition before applying the resonator metallization. In a final 
device this layer could, e.g., separate contact electrodes and nanotube from local top 
gate electrodes.

\begin{table}[t]
\begin{center}
 \begin{tabular}{rrr}
  $\AlO$ thickness & $\fr$ (GHz) & $\Qi$ ($4.2\un{K}$) \\ \hline
  $0\un{nm}$  & 3.510 & 2060 \\
  $0\un{nm}$  & 3.777 & 2170 \\
  $0\un{nm}$  & 4.265 & 1600 \\ \hline
  $10\un{nm}$ & 3.560 & 2690 \\
  $10\un{nm}$ & 3.840 & 2480 \\ \hline
  $150\un{nm}$& 3.576 & 2570 \\
  $150\un{nm}$& 3.842 & 1420 \\
  $150\un{nm}$& 4.350 & 1900 \\
 \end{tabular}
 \end{center}
 \caption{\label{t:qi}
 Internal quality factors \Qi\ at $T=4.2\un{K}$ of devices without or with 
 additional \AlO\ layer below pristine molybdenum-rhenium, cf. 
 Fig.~\ref{substrate}(e). At this temperature within experimental scatter no 
 impact of the \AlO\ on the quality factor can be seen.
 } 
\end{table}
Table~\ref{t:qi} shows an overview of internal quality factors \Qi\ at  
$T=4.2\un{K}$ of devices without or with additional \AlO\ layer below pristine 
molybdenum-rhenium. While there is a certain scatter, no clear tendency towards 
lower or higher quality factor is recognizable; within the experimental 
resolution achievable at this temperature no impact of the layer can be seen.

\section{Conclusion}

We demonstrate that \Lq\ coplanar waveguide resonators fabricated from a 
molybdenum-rhenium alloy display resonant behaviour at millikelvin temperatures 
even after being exposed to the chemical vapor deposition conditions required 
for carbon nanotube growth. Compared to a reference niobium device, the 
resonance frequency of lithographically identical molybdenum-rhenium devices is 
lower due to a larger kinetic inductance fraction of the material. The 
temperature dependence of resonance frequency and internal quality factor can 
be well described by Mattis-Bardeen theory in the analyzed temperature range.
 Both Si/\SiO\ and \AlO\ substrates can be used, with the expected impact of the 
permittivities on the resonance frequency. An additional \AlO\ layer deposited 
on a Si/\SiO\ substrate via atomic layer deposition, which could, e.g., be used 
for nanotube top gate isolation, does not lead to any change in quality factor
detectable at $T=4.2\un{K}$.

While our thin film deposition via co-sputtering from two sources has the 
advantage of adjustable alloy composition \cite{remo}, other published results 
using pre-alloyed sputter targets display significantly higher quality factors 
\cite{apl-singh:222601}. Given our excellent results using niobium, we can 
exclude problems with substrate, lithography, and detection circuitry. Further 
work shall thus target additional optimization of the metal thin films 
regarding both nominal composition and detailed deposition parameters.

We thank T. N. G. Meier and M. Kronseder for experimental help with XPS 
spectroscopy of the metal films. The authors gratefully acknowledge funding by 
the Deutsche Forschungsgemeinschaft via SFB 631, SFB 689, GRK 1570, and Emmy 
Noether project Hu 1808/1.

\bibliography{paper}

\end{document}